\documentclass{article}
\usepackage[utf8]{inputenc}
\usepackage{amsmath}
\usepackage{graphicx}
\usepackage{amssymb}
\DeclareMathOperator{\arcsinh}{arcsinh}
\usepackage{arxiv}
\usepackage{tikz}
\usepackage{pgfplots}
\usepackage{url}
\usetikzlibrary{plotmarks}
\usetikzlibrary{datavisualization} 
\usetikzlibrary{patterns}
\usetikzlibrary{shapes,arrows.meta,calc,fit,backgrounds,shapes.multipart,positioning}
\usetikzlibrary{backgrounds}

\title{Maximum transmission reach for optical signals in elastic optical networks employing band division multiplexing}

\author{
  Esteban Paz\\
  \texttt{epaz@udec.cl} \\
  Electrical Engineering Department\\
  Universidad de Concepcion
   \And
  Gabriel Saavedra \\
  \texttt{gasaavedra@udec.cl} \\
  Electrical Engineering Department\\
  Universidad de Concepcion
}

\begin{document}
\maketitle

\begin{abstract}
    Multi-band transmission systems have emerged as a potential answer to the limited capacity of silica based optical fiber. It is based on utilizing the complete low-loss region of optical fibers. Additionally, elastic optical networks (EON) were proposed to efficiently manage limited resources in optical networks. The selection of route, modulation format and spectrum is the main problem that requires solving to operate an EON. Understanding how linear and nonlinear impairments accumulate during propagation in an optical fiber link is essential to solve this problem. Here we preset a study on the maximum reach optical signals can propagate in a multi-band environment for a variety of modulation formats with commonly used bit error rate threshold values in EON.    
\end{abstract}

\section{Introduction}
Optical fibers underpin the global telecommunications infrastructure and carry over 95\% of the data traffic in the world. However, it has been acknowledged that the capacity of optical fibers is not infinite, and that nonlinear effects that occur while information is propagated through optical fibers impose a limit on the maximum achievable information rates of optical communication systems \cite{Mitra2001,Essiambre2010,Winzer20years,Ellis_2016}. To sustain the ever-increasing demand for data new or alternative solutions for optical communication systems are needed. Due to the large and unused available spectrum of optical fibers, adopting multiple transmission windows (multi-band) to transmit information can lead to capacity increases for optical communication systems. In fact, projections for the next 20 years have highlighted the potential of this solution to overcome the so-called capacity crunch imposed by the nonlinear nature of silica based optical fibers \cite{Winzer20years}. Although adopting this solution conveys many technological challenges in terms of optical devices and integration~\cite{Napoli18,NapoliMultiband} it is the most promising solution to maintain the thousand of kilometers of optical fiber infrastructure installed around the world. Despite this, the use of multiple transmission bands for transmission of signal enhances the nonlinear interactions between signals, that depend on the power of the transmitted signal and bandwidths~\cite{BayvelCapacity,SemrauOpEx,SaavedraGN}. One of the main challenges to tackle in this multi-band transmission regime is the presence of inter-channel stimulated Raman scattering (ISRS) that transfers energy between the co-propagating channels and dominates the system performance~\cite{SemrauOpEx,SemrauISRS,Cantono1,Cantono2}.

On the networking scenario, elastic optical networks (EONs) were proposed to overcome the limitations of WDM networks, using flexible resource allocation based on the demand of the users~\cite{Gerstel2012,Layec2013,Sambo2012,Velasco,Talebi2014}. In order to do this, the optical spectrum is divided into frequency slots units (FSUs) of 12.5 or 6.25 GHz, that are grouped together to provide the required transmission resources for a given user. This can lead to spectral savings compared to conventional WDM networks.
The main task a network operator must solve is to assign a path, a spectral region and a modulation format, known as the Routing, Modulation Level and Spectrum Assignment (RMLSA) problem \cite{Yuan2019b,Fontinele2017,Hashimoto2012,Luo2017,Klinkowski2011b,Christodoulopoulos2011}. To select the modulation format the physical impairments accumulated during propagation need to be considered, as they limit the maximum reach a signal can travel and be detected complying with the quality of service required by the user or network operator.   

Multi-band transmission, also known as band division multiplexing (BDM), can provide significant benefits to EONs operators, allowing them to support a larger number of users. However, in this scenario the RMLSA becomes a complex problem, as the nonlinear interactions arising from inter-channel stimulated Raman scattering introduce further physical limitations in the channel, translated into distinct performance for each transmission band. In this work we study the effect that linear and nonlinear noise accumulation in BDM systems has on the transmission reach for optical signals managed by an EON operator. Using fiber propagation models~\cite{SemrauISRS,SemrauClosed}  we estimate the maximum transmission reach for a variety of modulation formats commonly used in EONs for each transmission band in a BDM network.    

\section{Methodology}
To assign a wavelength and a modulation format for a user in EONs it is necessary to know the maximum transmission distance that an optical signal can travel to ensure a minimum quality of service for the user. The quality of service is typically ensured when the user exhibits a bit-error-rate (BER) lower than a predetermined threshold. Physical impairments, such as amplified spontaneous emission (ASE) noise, chromatic dispersion and nonlinear effects \cite{Politi2013} degrade the received signal-to-noise ratio (SNR), and thus, increase the BER of a transmission.
To achieve maximum transmission distance, the optical power launched into each fiber needs to be optimized to balance the effects of linear and nonlinear impairments, which have a strong dependence on the signal wavelength. To provide general conclusions, we compute a mean SNR for every transmission band considered in the system under study.

This section presents the methodology used to compute the per band SNR, and the parameters of the system under consideration.

\subsection{Transmission system model}
In this work the signal-to-noise ratio (SNR) was used as the performance metric to estimate the maximum transmission reach for each modulation format. The SNR was calculated as follows:
\begin{equation}
    SNR_{ch} = \frac{P_{ch}}{P_{ASE,ch}+\eta_{P_{ch}}P_{ch}^3},
    \label{eq:SNR}
\end{equation}
where $P_{ch}$ is the launch power of channel $ch$, $P_{ASE,ch}$  corresponds to the amplified spontaneous emission (ASE) noise power in the channel bandwidth, and $\eta_{P_{ch}}$ is the nonlinear interference (NLI) coefficient of said channel. The latter term ($\eta_{P_{ch}}P_{ch}^3$) in the denominator corresponds to the nonlinear "noise" generated during transmission, and it depends on the signal power and wavelength. 

Here we consider a system amplified by lumped optical amplifiers (such as EDFA, PDFA, NDFA and TDFA~\cite{Napoli18,NapoliMultiband}), then, the ASE noise contribution in Eq.~\eqref{eq:SNR} is given by:
\begin{equation}
    P_{ASE,ch} = 2n_{sp}hv(G-1),
    \label{eq:Pase}
\end{equation}
where $n_{sp}$ is the population inversion factor, $h$ is Planck's constant, $v$ the frequency of the channel of interest and $G = \alpha L$ is the amplifier gain, being $\alpha$ the attenuation coefficient of the fiber and $L$ the span length. Both the channel frequency and the amplifier gain vary significantly within the bandwidth of interest, and thus, the ASE noise is expected to vary in the same manner. 

The calculation of the nonlinear coefficients $\eta_{P_{ch}}$ was performed utilizing a closed-form approximation of the Gaussian noise model from \cite{SemrauClosed}. The nonlinear interference (NLI) coefficient $\eta$ is dominated by modified signal power profile from inter-channel stimulated Raman scattering (ISRS), self- and cross-phase modulations (SPM and XPM, respectively), and Four wave mixing (FWM). The NLI coefficient was obtained using:

\begin{equation}
    \eta(f_{i})_{ch,P} \approx \sum_{j=1}^{n} [\frac{P_{i,j}}{P_{i}}]^2\cdot[\eta_{SPM,j}(f_{i}) + \eta_{XPM,j}(f_{i})]
    \label{eq:eta}
\end{equation}
$$ $$
where $P_{i,j}$ is the power of channel $i$ launched into the $j-th$ span.
The contribution of SPM ($\eta_{SPM,j}$) is given by:
\begin{equation}
    \eta(f_{i})_{SPM} \approx  \frac{4}{9}\frac{\gamma^2}{B_{i}^2}\frac{\pi}{\phi_{i}\bar{\alpha}(2\alpha + \bar{\alpha})}\cdot[\frac{T_{i}-\alpha^2}{a}\arcsinh(\frac{\phi_{i}-B_{i}^2}{a\pi})+\frac{A^2-T_{i}}{A}\arcsinh(\frac{\phi_{i}-B_{i}^2}{A\pi})],
\end{equation}
$$ $$
with $ \phi_{i}=\frac{2}{3}\pi^2(\beta_{2}+2\pi\beta_{3}f_{i}), A=\alpha + \bar{\alpha}, T_{i}= (\alpha + \bar{\alpha}-P_{tot}C_{r}f_{i})^2.$

$B_i$ is the channel $i$ bandwidth, $\bar{\alpha}$ is considered equal to $\alpha$, $C_{r}$ is the Raman gain spectrum slope,$P_{tot}$ is the total launch power, $\beta_2$ is the group velocity dispersion and $\beta_3$ is the third-order dispersion parameter.

The XPM ($\eta_{XPM,j}$) contribution is given by:
\begin{equation}
    \eta(f_{i})_{XPM} \approx  \frac{32}{27}\sum_{k=1,k\neq i}^{N_{ch}}(\frac{P_{k}}{P_{i}})^2 \frac{\gamma^2}{B_{k}\phi_{i,k}\bar{\alpha}(2\alpha+\bar{\alpha})}\cdot[\frac{T_{k}-\alpha^2}{\alpha}\arctan{(\frac{\phi_{i,k}-B_{i}^2}{\alpha})}+\frac{A^2-T_{k}}{A}\arctan{(\frac{\phi_{i}-B_{i}^2}{A}})],
\end{equation}
$$ $$
with $\phi_{i,k} = 2\pi^2(f_{k}-f_{i})[\beta_{2}+\pi\beta_{3}(f_{k}+f_{i})]$, as is shown in [1].

Further, to relate the received SNR to a maximum transmission reach, BER thresholds conventionally used in coherent transmission systems and EONs were used. The relationship between SNR and BER, assuming an additive white Gaussian noise channel, is given by~\cite{salehi2007digital}:

\begin{equation}
    {BER }_{\Lambda-PSK}=\frac{1}{2}\mathrm{erfc}\left( \sqrt{\frac{SNR}{\lambda}\frac{{ \Delta }_{ref}}{{B}_{ch}}}\right),
\label{eq:BER_PSK}
\end{equation}

\begin{equation}
\begin{split}
    { BER }_{\Lambda-QAM}=&\frac{1}{\lambda}\left(1-\frac{1}{\sqrt{\Lambda}}\right)
    \cdot\mathrm{erfc}\left( \sqrt{\frac{3\ SNR}{2(\Lambda-1)}\frac{{ \Delta }_{ref}}{{B}_{ch}}}\right),
\label{eq:BER_QAM}
\end{split}
\end{equation}
where  $\Lambda$ is the constellation of cardinality, $\lambda$ represents the bits encoded in each constellation symbol, such that $ \Lambda=2^\lambda$. $\Delta_{ref}$ and $B_{ch}$ represent the reference and channel bandwith.

\subsection{System Parameters}

Here we consider that the transmission link is based on standard single mode fiber (SMF) and the coherent system employs BDM. A total of 2720 FSUs were occupied centered in the S band, yielding a total bandwidth of 34THz, from 1365-1615 nm. The bandwidth was selected to cover the E, S, C and L bands. The O band was not considered due to the enhancement of nonlinear interactions as the zero dispersion wavelengths is approached, and the reduced accuracy of the GN model in that region. The Raman gain coefficient ($C_r$) was approximated assuming a linear increase for frequency shifts between 0-13 THz, for larger frequency shifts the value was assumed to be 0.  
Further, we assume that the signal launch power is the same in every transmission band.

The simulation parameters are shown in table \ref{tab:parameters}. 

\begin{table}[h!]
\centering
\begin{tabular}{|l |c |c|}
\cline{1-3}
Parameter                   & Symbol  & Value \\
\hline
\multicolumn{3}{|l|}{\textit{Fiber parameters}}  \\
\hline
Attenuation Coefficient @ 1550 nm  & $\alpha$ & $\mathrm{0.165 dB/km}$\\
Attenuation Coefficient @ 1590 nm  & $\alpha$  & $\mathrm{0.171 dB/km}$\\
Attenuation Coefficient @ 1495 nm  & $\alpha$  & $\mathrm{0.177 dB/km}$\\
Attenuation Coefficient @ 1410 nm & $\alpha$  & $\mathrm{0.217 dB/km}$\\
Dispersion Coefficient & $D$ &  $\mathrm{21.3~{ps}/{nm}/{km}}$\\
Dispersion Slope            & $S$ & $\mathrm{0.067~{ps}/{nm}^{2}/{km}}$ \\
Raman gain coefficient @ 13 THz          &$C_r$& $\mathrm{0.028}$ $\mathrm{(1/W/km)}$\\
Nonlinear coefficient & $\gamma$ & $\mathrm{1.2~ W^{-1}km^{-1}}$ \\
Optical Reference Bandwidth & ${\Delta}_{ref}$ & $\mathrm{12.5 GHz}$ \\ 
Span length  & $L$ & $\mathrm{100 km}$\\
\hline
\multicolumn{3}{|l|}{\textit{Amplifier parameters}}  \\
\hline
Noise figure & ${F}$ & $\mathrm{5 dB} $\\
\hline
\multicolumn{3}{|l|}{\textit{Signal parameters}}  \\
\hline
Central frequency WDM & $v$ &  $\mathrm{1480 nm (200.67 THz)}$\\
Symbol Rate & $R$ & $\mathrm{12.5 GBaud}$ \\
Channel spacing & $\Delta f$ &  $\mathrm{12.5 GHz}$\\ 
Optical bandwidth & BW & $\mathrm{2720 FSUs (34 THz)}$\\
\hline
\end{tabular}
\caption{Fiber parameters for maximum reach calculation} \label{tab:parameters}
\end{table}

\section{Results}
\subsection{Received SNR}
\begin{figure}[h!]
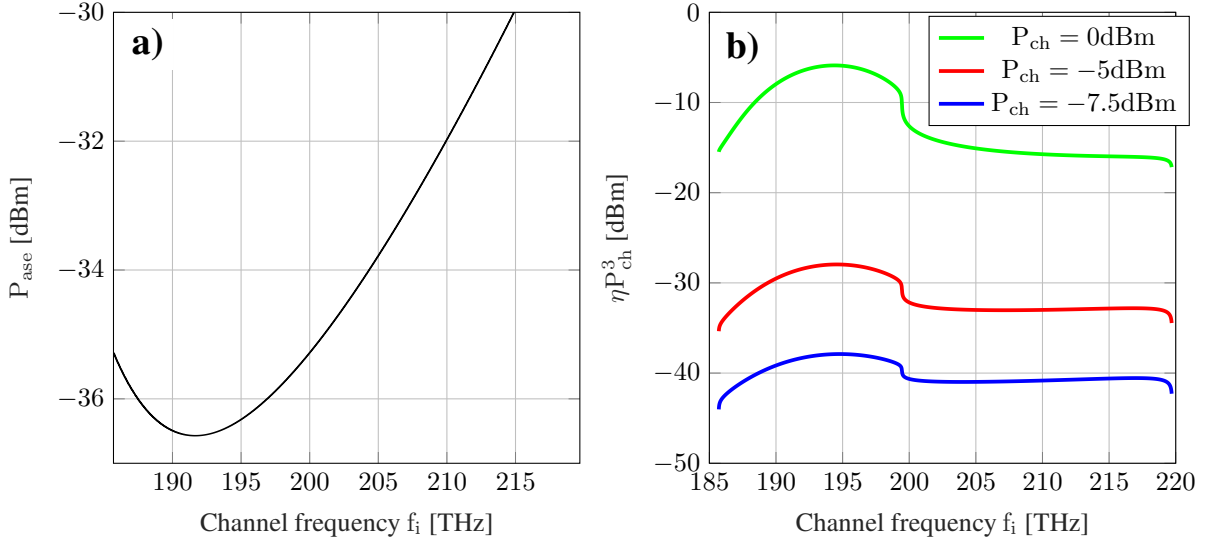

    \centering
  \input{Figures/Pase_chfreq}
    \input{Figures/Eta_chfreq}
    \caption{\textbf{a)} ASE noise power as a function of channel frequency. Solid black line represents ASE noise for every channel under study,colored lines show the averge ASE in every transmission band. \textbf{b)} Nonlinear Interference noise $\mathrm{\eta P_{ase}^3}$ as a function of the channels frequency, for $\mathrm{P_{launch} = 0dBm, -5dBm, -7.5dBm}$.}
    \label{fig:Noise_chfreq}
\end{figure}

First we study the noise generation over the entire bandwidth of interest in a single fiber span. To do so, Eq.~\eqref{eq:Pase} and Eq.~\eqref{eq:eta} were used to calculate the ASE and NLI coefficient, respectively. The ASE noise power arising from optical amplification ($P_{ase,ch}$) as a function of the channel frequency is presented in Fig.~\ref{fig:Noise_chfreq} \textbf{a)}. The $P_{ase,ch}$ for each channel within the studied bandwidth is shown in solid black. A strong dependency on the channel frequency is observed, both due to the attenuation coefficient and the nature of Eq.~\eqref{eq:Pase}. Remark that $P_{ase}$ within the E band is on average $6dB$ higher than the noise power of the C band. This agrees with theory, since the frequencies in the E band present a higher attenuation coefficient, and consequently a higher gain is needed for compensation.

Secondly, the NLI coefficient for each FSU was obtained for the system under study. The values of $\eta $ from Eq.~\eqref{eq:eta} were computed as a function of the channel frequency at a variety of launch powers, from $\mathrm{-25}$ to $\mathrm{10}$ dBm. In Figure ~\ref{fig:Noise_chfreq} \textbf{b)} particular plots for NLI at $\mathrm{P_{launch}}$ =-7.5 dBm, -5 dBm, 0 dBm are presented. 
An increase in NLI noise is observed for the lower frequencies of the spectrum, corresponding to the region where the channels experience amplification due to the power transfer arising from ISRS. Additionally, it can be observed that a lower launch power translates into a lower value of NLI noise.   

\begin{figure}[h!]
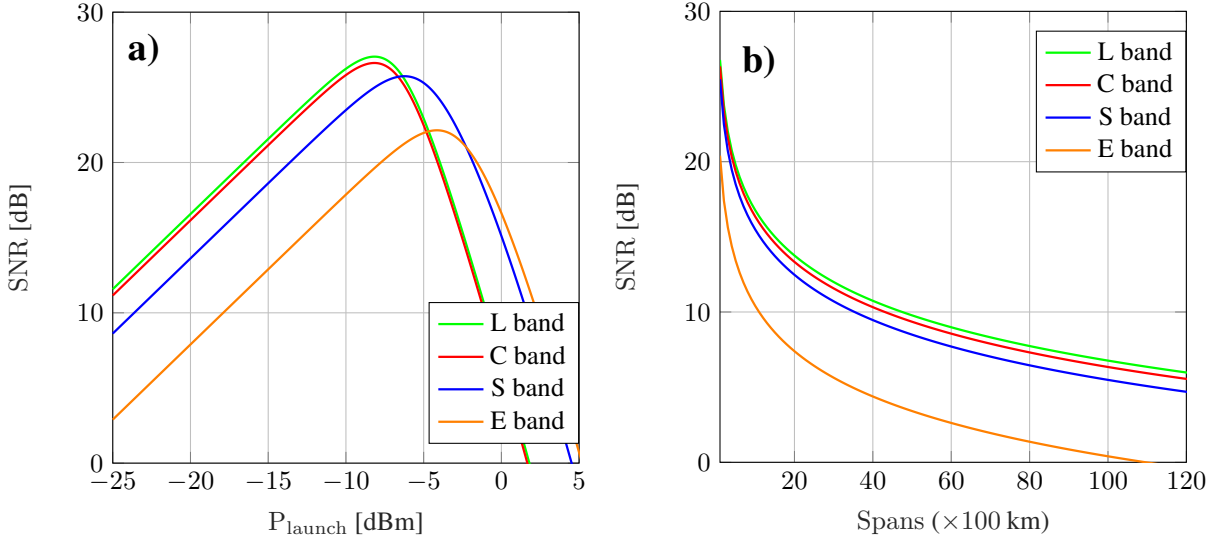

    \centering
  \input{Figures/SNR_Plaunch_2}
    \input{Figures/SNR_span_2}
    \caption{\textbf{a)}SNR as a function of the Launch Power $\mathrm{P_{launch}}$ in dBm. \textbf{b)}SNR in function of the number of spans}
    \label{fig:SNR}
\end{figure}

 The SNR after one fiber span was calculated as a function of the signal launch power using Eq.~\eqref{eq:SNR}. To estimate the maximum reach for each transmission band the FSU with the lowest SNR was selected to represent each band. This ensures that every channel within the band can, at least, propagate the estimated distance.
 The results are shown in Fig. ~\ref{fig:SNR} \textbf{a)} from which is observable that the L band takes a maximum $\mathrm{27.1}$ dB SNR at $\mathrm{-7.5}$ dBm, followed by the C band with SNR=$\mathrm{26.8}$ dB at $\mathrm{-8}$ dBm. The worst maximum SNR occurs on the E band with a value of $\mathrm{23.8}$ dB at $\mathrm{-5.5}$ dBm. 
From Figure ~\ref{fig:Noise_chfreq} \textbf{b)} is observed that the lowest NLI noise at $P_{launch}$ = -7.5 dBm happens around $185-190Thz$ and $205-215THz$, corresponding to the L band and part of the C and S bands respectively. Remark that the lowest ASE noise power in Figure ~\ref{fig:Noise_chfreq} \textbf{a)} is found in the L and S bands, so the resulting SNRs in L and C bands are expected. The resulting SNR values are summarized in Table ~\ref{tab:SNR1}. Which indicate the maximum SNR obtained paired the optimal launch power for each band.

\begin{table}[h!]
\centering
\begin{tabular}{|l |c|c|}
\cline{1-3}
 Band                     & SNR[dB]   & $P_{launch}$ [dBm] \\
\hline
 E (1365nm - 1460nm) &   23.8        &   -5.5  \\
 S (1460nm - 1530nm) &   26.3       &    -7 \\
 C (1530nm - 1565nm) &   26.8        &   -8  \\
 L (1565nm - 1615nm) &   27.1        &  -7.5   \\

\hline
\end{tabular}
\caption{ Maximum SNR per band and its respective launch power} \label{tab:SNR1}
\end{table}

Now, Fig.~\ref{fig:SNR} \textbf{b)} presents the SNR for each transmission band as a function of the number of spans. This was obtained assuming incoherent addition of NLI as follows:

\begin{equation}
    SNR_{ch} = \frac{P_{ch}}{N P_{ASE,ch}+N \eta_{ch}P_{ch}^3},
    \label{eq:SNR_spans}
\end{equation}
where N is the number of transmission spans.

In these calculations a signal launch power of $\mathrm{P_{launch}}$ =- 7.5 dBm was considered for every band, and consequently the starting SNR for each band might be slightly lower than the optimum SNR. Considering that the L band has the higher SNR among the studied bands at $\mathrm{-6.5}$ dBm it presents the highest SNR throughout all the spans, since a proportional decay is observed for every transmission band. In the same manner, the E band presents the lowest SNR as a function of the number of spans. 

\subsection{Maximum transmission reach}

The results presented so far show the generation and accumulation of noise for a optical signal propagating in a multi band transmission system. A network operator uses this information to estimate the maximum a distance can be transmitted and detected with a given quality of service. Here we provide the bit rates achieved by signal in a single FSU over 2 polarizations with and without FEC overhead for a variety of modulation formats of interest. Then we obtain the maximum transmission reach for said signals. The modulation formats considered here are: BPSK, QPSK, 16QAM ,64QAM and 256QAM; the BER thresholds ($\mathrm{BER_{th}}$) considered are: $4.7\times10^{-3}$ \cite{Staircodes},$10^{-6}$,$10^{-9}$ \cite{fallahpour2014energy,beyranvand2013quality,aibin2015adaptive}.    
The bit rates achieved in a single FSU are shown in Table~\ref{tab:Rate_ModFormat}.

\begin{table}[h!]
\centering
\begin{tabular}{|l |c|c|}
\cline{1-3}
          & $\mathrm{BER_{th}=4.7\times 10^{-3}}$ & $\mathrm{BER_{th}=10^{-6} \& 10^{-9}}$\\
\hline
Modulation format & Net bit rate [Gb/s]  & Net bit rate [Gb/s]\\
\hline
 BPSK          & 23 & 25\\
 QPSK          & 46 & 50\\
 8QAM          & 69 & 75\\
 16QAM         & 92 & 100\\
32QAM          & 115 & 125\\
 64QAM         & 140 & 150\\
 256QAM        & 186 & 200\\
\hline
\end{tabular}
\caption{Net bit rate for each modulation formation format using a single FSU} 
\label{tab:Rate_ModFormat}
\end{table}

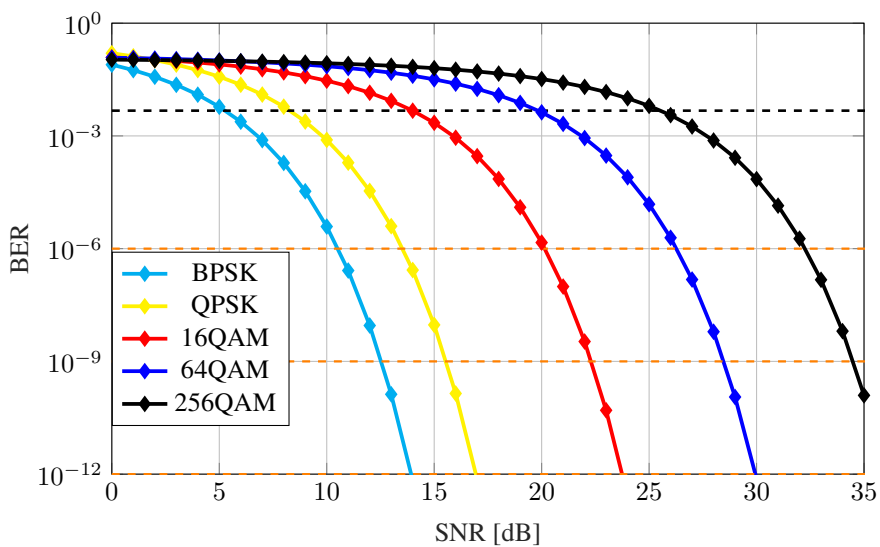
\begin{figure}[h!]
    \centering
  \begin{tikzpicture}
\begin{axis}[
    width=10cm,
    height=6cm,
at={(0.769in,0.47in)},
scale only axis,
xmin=0,
xmax=35,
xlabel style={font=\color{white!15!black}},
xlabel={SNR [dB]},
ymode=log,
ymin=1e-12,
ymax=1,
ylabel style={font=\color{white!15!black}},
ylabel={BER},
axis background/.style={fill=white},
xmajorgrids,
ymajorgrids,
legend style={at={(0,0.3)},anchor=west},
]

\addplot[color=cyan,mark=diamond*, mark options={solid, fill=cyan},line width=0.5mm]
  table[row sep=crcr]{
0 0.0786496035251426\\
1 0.0562819519765415\\
2 0.0375061283589260\\
3 0.0228784075610853\\
4 0.0125008180407376\\
5 0.00595386714777866\\
6 0.00238829078093281\\
7 0.000772674815378444\\
8 0.000190907774075993\\
9 3.36272284196175e-05\\
10 3.87210821552204e-06\\
11 2.61306795357520e-07\\
12 9.00601035062875e-09\\
13 1.33293101753005e-10\\
14 6.81018912878076e-13\\
15 9.12395736262809e-16\\
16 2.26739584445441e-19\\
17 6.75896977065469e-24\\
18 1.39601431090675e-29\\
19 1.00107397357086e-36\\
20 1.04424379188127e-45\\
21 5.29969722887031e-57\\
22 3.29608811928451e-71\\
23 4.42763849843474e-89\\
24 1.44441906577261e-111\\
25 7.30696918464799e-140\\
26 1.79516322712013e-175\\
27 2.73597996566879e-220\\
28 1.06846073755641e-276\\
29 0\\
30 0\\
31 0\\
32 0\\
33 0\\
34 0\\
35 0\\};

\addplot[color=yellow,mark=diamond*, mark options={solid, fill=yellow},line width=0.5mm]
  table[row sep=crcr]{
0 0.158655253931457\\
1 0.130927296755524\\
2 0.104028637085389\\
3 0.0788958719817244\\
4 0.0564953017493617\\
5 0.0376789881474634\\
6 0.0230071388778660\\
7 0.0125870331221446\\
8 0.00600438640016356\\
9 0.00241331041963386\\
10 0.000782701129001274\\
11 0.000193985472057861\\
12 3.43026238664154e-05\\
13 3.96924839634283e-06\\
14 2.69514811736672e-07\\
15 9.36103999068510e-09\\
16 1.39902780939769e-10\\
17 7.23597570853658e-13\\
18 9.84500243879789e-16\\
19 2.49451671785946e-19\\
20 7.61985302416050e-24\\
21 1.62296192653899e-29\\
22 1.20974376996816e-36\\
23 1.32491919348297e-45\\
24 7.14952466598120e-57\\
25 4.80352986222599e-71\\
26 7.11135314237203e-89\\
27 2.62194050326119e-111\\
28 1.54731219917564e-139\\
29 4.61509293895264e-175\\
30 8.97916392400319e-220\\
31 4.76854670722285e-276\\
32 0\\
33 0\\
34 0\\
35 0\\};


\addplot[color=red,mark=diamond*, mark options={solid, fill=red},line width=0.5mm]
  table[row sep=crcr]{
0  0.122760158628483\\
1 0.115466426518155\\
2 0.107517947843041\\
3 0.0989210644710998\\ 
4 0.0897107665378770\\
5 0.0799605068841076\\
6 0.0697926815562709\\
7 0.0593881195535992\\ 
8 0.0489916831960609\\
9 	0.0389096609825274\\
10 0.0294936013219285\\ 
11 	0.0211057319912030\\
12 0.0140647981345973\\
13 0.00857940283540700\\ 
14 0.00468780676527659\\
15 0.00223270018041700\\
16 0.000895609042849801\\ 
17 0.000289753055766917\\ 
18 7.15904152784974e-05\\
19 1.26102106573566e-05\\
20 1.45204058082076e-06\\
21 9.79900482590699e-08\\
22 3.37725388148575e-09\\
23 4.99849131573773e-11\\
24 2.55382092329279e-13\\
25 3.42148401098557e-16\\
26 8.50273441670405e-20\\
27 2.53461366399545e-24\\
28 5.23505366590046e-30\\
29 3.75402740089060e-37\\
30 3.91591421955477e-46\\
31 1.98738646082637e-57\\
32 1.23603304473169e-71\\
32 1.66036443691303e-89\\
33 5.41657149664727e-112\\
34 2.74011344424299e-140\\
35 0\\};


\addplot[color=blue,mark=diamond*, mark options={solid, fill=blue},line width=0.5mm]
  table[row sep=crcr]{
  0 0.120641988040395\\ 1 0.117625788644167\\ 2 0.114264985335776\\ 3 0.110526935312723\\
  4 0.106378694747346\\
  5 0.101788351927295\\
  6 0.0967269781989794\\
  7 0.0911713814483950\\
  8 0.0851078618670935\\
  9 0.0785371520058295\\
  10 0.0714806400606411\\
  11 0.0639877776918873\\
  12 0.0561441960619269\\
  13 0.0480794276821975\\
  14 0.0399722251136646\\
  15 0.0320503703986822\\
  16 0.0245809342731719\\
  17 0.0178469487304694\\
  18 0.0121086263547943\\
  19 0.00755281736524975\\
  20 0.00424321504559928\\
  21 0.00209233402659440\\
  22 0.000876551410118952\\
  23 0.000299442597190263\\
  24 7.92095153558772e-05\\
  25 1.52004620046546e-05\\	
  26 1.94926541216136e-06\\	
  27 1.50611869857474e-07\\	
  28 6.15427786240665e-09\\	
  29 1.12842343114206e-10\\
  30 7.54878404144402e-13\\
  31 1.41977352958212e-15\\
  32 5.40740855914019e-19\\
  33 2.75898748190030e-23\\
  34 1.12092566440710e-28\\
  35 1.88376817857561e-35\\
};
\addplot[color=black,mark=diamond*, mark options={solid, fill=black},line width=0.5mm]
  table[row sep=crcr]{
  0 0.107065624343304\\
  1 0.105836322932501\\
  2 0.104459389514190\\
  3 0.102917779596656\\
  4 0.101192763801010\\	
  5 0.0992638812811604\\
  6 0.0971089500716120\\
  7 0.0947041648633554\\
  8 0.0920243244338240\\
  9 0.0890432461527636\\
  10 0.0857344440187384\\
  11 0.0820721693400485\\
  12 0.0780329379925491\\
  13 0.0735976912862160\\
  14 0.0687547507239143\\
  15 0.0635037152999646\\
  16 0.0578603881735190\\
  17 0.0518626689861829\\
  18 0.0455770569894065\\
  19 0.0391049229084234\\
  20 0.0325869945258850\\
  21 0.0262036241795384\\
  22 0.0201676295963604\\
  23 0.0147064194121552\\
  24 0.0100316960281937\\
  25 0.00629927973817447\\
  26 0.00356854970547827\\
  27 0.00177803321441000\\
  28 0.000754621562336096\\
  29 0.000262019543341213\\
  30 7.07395544687615e-05\\
  31 1.39274292895061e-05\\
  32 1.84445422809404e-06\\
  33 1.48400110690408e-07\\
  34 6.38057183214595e-09\\
  35 1.24728913630848e-10\\
};

\addplot[dashed, color=orange,line width=0.3mm]
  table[row sep=crcr]{
  0  1e-6\\
  36 1e-6\\
};

\addplot[dashed, color=orange,line width=0.3mm]
  table[row sep=crcr]{
  0  1e-9\\
  36 1e-9\\
};
\addplot[dashed, color=orange,line width=0.3mm]
  table[row sep=crcr]{
  0  1e-12\\
  36 1e-12\\
};
\addplot[dashed, color=black,line width=0.35mm]
  table[row sep=crcr]{
  0  4.7e-3\\
  36 4.7e-3\\
};

\legend{BPSK,QPSK,16QAM,64QAM,256QAM}
\end{axis}
\end{tikzpicture}
    \caption{Relationship between SNR and BER for different M-PSK and M-QAM modulation formats}
    \label{fig:BER_SNR}
\end{figure}

The relationship between the SNR and BER is shown in Fig.~\ref{fig:BER_SNR}, results obtained using Eqs.~\eqref{eq:BER_QAM} and~\eqref{eq:BER_PSK}.

For BPSK, which is the less complex modulation format present, one can notice that a lower SNR is required in comparison to more complex formats for a fixed BER value. In general, as the modulation formats are able to transmit more bits in each symbol a higher SNR is required to obtain a fixed BER. Additionally, the studied BER thresholds are shown with dashed horizontal lines.

The maximum reach the modulations formats studied can achieve for the BER thresholds under considerations are summarized in Tables~\ref{tab:reach_FEC},~\ref{tab:reach_e-6},~\ref{tab:reach_e-9}. These values together with the ones presented in Table~\ref{tab:Rate_ModFormat} can be used by an EON operator as the input to solve the RMLSA problem. 

\begin{table}[h!]
\centering
\begin{tabular}{|l |c|c|c|c|c|}
\cline{1-6}
Modulation format           & BPSK & QPSK &16QAM&64QAM & 256QAM \\
\hline
SNR threshold  & 5.3 dB & 8.3 dB& 14 dB& 19.8 dB& 25.5 dB\\
\hline
Band                 & Reach[spans] &Reach[spans]& Reach[spans] & Reach[spans] & Reach [spans] \\
\hline
 E          & 31 & 15& 4 & 1 & 0\\
 S          & 102 & 51& 14 & 3 & 0\\
 C          & 130 & 65& 17 & 4 & 1\\
 L          & 144 & 72& 19 & 5 & 1\\
\hline
\end{tabular}
\caption{Maximum reach for a BER threshold of $4.7\times10^{-3}$} 
\label{tab:reach_FEC}
\end{table}

\begin{table}[h!]
\centering
\begin{tabular}{|l |c|c|c|c|c|}
\cline{1-6}
Modulation format           & BPSK & QPSK &16QAM&64QAM & 256QAM \\
\hline
SNR threshold  & 10.5 dB & 13.5 dB& 20.2 dB& 26.3 dB& 32.3 dB\\
\hline
Band                 & Reach[spans] &Reach[spans]& Reach[spans] & Reach[spans] & Reach [spans] \\
\hline
 E          & 9 & 4& 1 & 0& 0\\
 S          & 31 & 15& 3 & 0 & 0\\
 C          & 39 & 19& 4 & 1 & 0\\
 L          & 43 & 22& 4 & 1 & 0\\
\hline
\end{tabular}
\caption{Maximum reach for a BER threshold of $10^{-6}$}
\label{tab:reach_e-6}
\end{table}

\begin{table}[h!]
\centering
\begin{tabular}{|l |c|c|c|c|c|}
\cline{1-6}
Modulation format           & BPSK & QPSK &16QAM&64QAM & 256QAM \\
\hline
SNR threshold  & 12.5dB & 15.6 dB& 22.3 dB& 28.5 dB& 34.6 dB\\
\hline
Band                 & Reach[spans] &Reach[spans]& Reach[spans] & Reach[spans] & Reach [spans] \\
\hline
 E          & 5 & 2& 0 & 0 & 0\\
 S          & 19 & 9& 2 & 0 & 0\\
 C          & 24 & 12& 2 & 0 & 0\\
L          & 27 & 13& 2 & 0 & 0\\
\hline
\end{tabular}
\caption{Maximum reach for a BER threshold of $10^{-9}$}
\label{tab:reach_e-9}
\end{table}

\section*{Conclusion}
In this work we computed the maximum reach for a variety of modulation formats with commonly used threshold values in EONs. This was achieved by computing the particular SNR for the E,S,C and L bands using a closed form approximation of the Gaussian noise model in presence of ISRS, which allowed us to obtain the nonlinear coefficients needed for the SNR calculation.
\section*{Acknowledgements}
The authors thank A. Lozada (USM) for helpful feedback and discussions.
\bibliographystyle{IEEEtran}
\bibliography{IEEEabrv,bibliography}
\end{document}